\documentclass[]{aa}  
\newcommand*{\aas}{\ensuremath{A\&A}}

\newcommand*{\fesc}{\ensuremath{f_{esc}}}

\newcommand*{\Ha}{H\ensuremath{\alpha}}
\newcommand*{\Hb}{H\ensuremath{\beta}}
\newcommand*{\Hg}{H\ensuremath{\gamma}}

\newcommand*{\xOII}{[O\,{\scshape ii}]\,\ensuremath{\lambda3727 \AA }}
\newcommand*{\xNeIII}{[Ne\,{\scshape iii}]\,\ensuremath{\lambda3868 \AA }}
\newcommand*{\xOIII}{[O\,{\scshape iii}]\,\ensuremath{\lambda5007 \ \AA \ }}
\newcommand*{\xOIIIuv}{[O\,{\scshape iii}]\,\ensuremath{\lambda1663 \AA}}

\newcommand*{\xCIII}{C\,{\scshape iii}\,\ensuremath{\lambda\lambda1907\text{-}1909}}

\newcommand*{\xHeII}{He\,{\scshape ii}\,\ensuremath{\lambda1640}}

\newcommand*{\CIII}{C\,{\scshape iii]}}

\newcommand*{\HeII}{He\,{\scshape ii}}
\newcommand*{\OIII}{[O\,{\scshape iii}]}

\usepackage{graphicx}

\usepackage[colorlinks=true,
            linkcolor=red,
            urlcolor=red,
            citecolor=blue]{hyperref}

\usepackage{txfonts}

\usepackage{multirow,array}

\usepackage{tikz}

\definecolor{lime}{HTML}{A6CE39}
\DeclareRobustCommand{\orcidicon}{%
    \begin{tikzpicture}
    \draw[lime, fill=lime] (0,0) 
    circle [radius=0.16] 
    node[white] {{\fontfamily{qag}\selectfont \tiny ID}};
    \draw[white, fill=white] (-0.0625,0.095) 
    circle [radius=0.007];
    \end{tikzpicture}
    \hspace{-2mm}
}

\foreach \x in {A, ..., Z}{%
    \expandafter\xdef\csname orcid\x\endcsname{\noexpand\href{https://orcid.org/\csname orcidauthor\x\endcsname}{\noexpand\orcidicon}}
}

\usepackage[rightcaption]{sidecap}

\begin{document}



\title{Evolution of the UV slope of galaxies at cosmic morning ($z>4$): the properties of extremely blue galaxies} 


\author{D. Dottorini\inst{1,2},
A. Calabr{\`o}\orcidA{}\inst{1},
L. Pentericci\inst{1},
S.Mascia\inst{1,3},
M. Llerena\inst{1},
L. Napolitano\inst{1,2},
P. Santini\inst{1},
G. Roberts-Borsani\inst{4},
M. Castellano\inst{1},
R. Amorin\inst{5,6},
M. Dickinson\inst{7}, 
A. Fontana\inst{1},
N. Hathi\inst{8},
M. Hirschmann\inst{9},
A. Koekemoer\inst{8}, 
R.A. Lucas\inst{8}, 
E. Merlin\inst{1},
A. Morales\inst{10},  
F. Pacucci\inst{11,12},
S. Wilkins\inst{13,14},
P. Arrabal Haro\thanks{NASA Postdoctoral Fellow}\inst{15}, 
M. Bagley\inst{10}, 
S. Finkelstein\inst{10}, 
J. Kartaltepe\inst{16}, 
C. Papovich\inst{17,18}, 
N. Pirzkal\inst{19} 
}

\institute{INAF - Osservatorio Astronomico di Roma, via Frascati 33, 00078, Monte Porzio Catone, Italy 
\and Universit\`a di Roma 'La Sapienza', piazzale Aldo Moro 5, 00185, Roma, Italy  
\and Institute of Science and Technology Austria (ISTA), Am Campus 1, A-3400 Klosterneuburg, Austria 
\and Department of Astronomy, University of Geneva, Chemin Pegasi 51, 1290 Versoix, Switzerland 
\and Instituto de Astrof\'isica de Andaluc\'ia (CSIC), Apartado 3004, 18080
Granada, Spain 
\and ARAID Foundation, Centro de Estudios de Física del Cosmos de
Arag\'on (CEFCA), Unidad Asociada al CSIC, Plaza San Juan 1, E–
44001 Teruel, Spain 
\and NSF’s National Optical-Infrared Astronomy Research Laboratory, 950 N. Cherry Ave., Tucson, AZ 85719, USA 
\and Space Telescope Science Institute, 3700 San Martin Drive, Baltimore, MD 21218, USA 
\and Institute of Physics, Laboratory of Galaxy Evolution, Ecole Polytechnique F\'ed\'erale de Lausanne (EPFL), Observatoire de Sauverny,
1290 Versoix, Switzerland 
\and Department of Astronomy, The University of Texas at Austin, 2515 Speedway, Austin, TX 78712, USA 
\and Center for Astrophysics | Harvard \& Smithsonian, 60 Garden St., Cambridge MA, 02138, USA 
\and Black Hole Initiative at Harvard University, 20 Garden St., Cambridge MA, 02138, USA 
\and Astronomy Centre, University of Sussex, Falmer, Brighton BN1 9QH, UK 
\and Institute of Space Sciences and Astronomy, University of Malta, Msida MSD 2080, Malta 
\and Astrophysics Science Division, NASA Goddard Space Flight Center, 8800 Greenbelt Rd, Greenbelt, MD 20771, USA 
\and Laboratory for Multiwavelength Astrophysics, School of Physics and Astronomy, Rochester Institute of Technology, 84 Lomb Memorial Drive, Rochester, NY 14623, USA  
\and George P.\ and Cynthia Woods Mitchell Institute for Fundamental Physics and Astronomy, Texas A\& M University, College Station, TX, 77843-4242 USA 
\and Department of Physics and Astronomy, Texas A\&M University, College Station, TX, 77843-4242, USA 
\and ESA/AURA, Space Telescope Science Institute, 3800 San Martin Drive, Baltimore, MD, 21218, USA 
}
\date{Submitted to A\&A}

\abstract {
We present an analysis of the ultraviolet (UV) continuum slope, $\beta$, using a sample of 733 galaxies selected from a mixture of JWST ERS/GTO/GO observational programs and with $z > 4$. We considered only spectroscopic data obtained with the low resolution ($R \sim 30-300$) PRISM/CLEAR NIRSpec configuration. 
Studying the correlation of $\beta$ with $M_{UV}$ we found an overall decreasing trend of $\beta = (-0.056\pm0.017)M_{UV}+(-3.01\pm0.34)$, consistent with brighter galaxies presenting redder $\beta$ values as found in previous works. However, analyzing the trend in separate redshift bins, we find that at high redshift the relation becomes much flatter and it is consistent with a flat slope (within $1 \sigma$).
Furthermore, we find that $\beta$ tends to decrease with redshift with an evolution as $\beta = (-0.075\pm0.010) z + (-1.496 \pm 0.056)$, consistent with most previous results that show a steepening of the spectra going at higher $z$. 
We then select a sample of galaxies with extremely blue slopes ($\beta < -2.6$): such slopes are steeper than what is predicted by stellar evolution models, even for dust free, young, metal poor populations, when the contribution of nebular emission is included. 
We select 51 extremely blue galaxies (XBGs) and we investigate the possible physical origin of their steep slopes, comparing them to a sub-sample of redder galaxies (matched in $\Delta z = \pm 0.5$ and $\Delta M_{UV} = \pm 0.2$). We find  that XBGs have younger stellar populations, stronger ionization fields, lower dust attenuation, and lower but not pristine metallicity ($\sim 10\% Z_\odot$) compared to red galaxies. However, these properties alone cannot explain the extreme $\beta$ values. By using indirect inference  of Lyman continuum escape, using the most recent models, we estimate values for the escape fraction $f_{esc} > 10\%$ in at least $25\%$ of the XBGs, while all the red sources have much smaller $f_{esc}$. A reduced nebular continuum contribution as due to either a high escape fraction or to a bursty star-formation history is likely the origin of the extremely blue slopes. 
}
%


\titlerunning{\footnotesize Spectroscopic UV slope of galaxies from redshift $4$}
\authorrunning{Dottorini et al.}
\maketitle

\section{Introduction}\label{sec:introduction}
 
Recent observations from JWST, thanks to its unprecedented sensitivity and resolution, have been able to probe the most distant galaxies ever observed, offering a direct glimpse into the first stages of galaxy formation and evolution \citep[e.g.][]{treu22,finkelstein23,adamo24}. These observations have also been fundamental to studying the process of cosmic reionization  between $\sim 300$ million and 1 billion years after the Big Bang (i.e., approximately between redshift $14$ and $6$), during which the ultraviolet (UV) light emitted by the first stars and galaxies reionized the surrounding neutral hydrogen in the intergalactic medium (IGM) \citep{jones24,napolitano24}. 

A critical observational diagnostic to understand early galaxy evolution and the reionization mechanism is the slope of the UV continuum $\beta$. The UV slope characterizes the spectral energy distribution (SED) of a galaxy in the UV range, which is dominated by the emission of massive O/B-type main sequence stars. Defined by the relation $f_\lambda \propto \lambda^\beta$ \citep{calzetti94} where $f_\lambda$ is the flux density at wavelength $\lambda$, $\beta$ provides essential insights into the physical properties of galaxies, including the dust content and the stellar population properties. Indeed, $\beta$ is primarily sensitive to the presence of dust, which can redden the SED and result in a less negative or even positive UV slope \citep{meurer99,calzetti00,wilkins16}. It is also directly influenced by the age and metallicity of the stellar populations within a galaxy (see e.g., \citealt{bouwens12,castellano12,tacchella22}); younger and more metal-poor galaxies tend to have bluer UV slopes, reflecting the predominance of hot, massive stars. 
Even though massive Pop III stars could have a very blue intrinsic UV spectrum ($\beta < -2.6$), their strong nebular continuum emission produces spectral shapes that are redder than non-Pop III stellar populations \citep{raiter10,zackrisson11,trussler23}.  
Furthermore, $\beta$ is also a crucial quantity to study reionization, since it is related to the fraction of Lyman continuum photons escaping from a galaxy f$_{\rm esc}$ \citep{mascia23,chisholm22}, whose direct estimation becomes prohibitive at $z>4$ due to the increasingly high opacity to the ionizing radiation of the IGM.   

The extensive imaging campaigns conducted by recent surveys with JWST NIRCam \citep{rieke05, rieke23} have made it possible to assemble a large statistical sample of photometrically selected galaxies at high redshifts ($\geq 4$) and to directly measure their UV slope from the photometry. Recent studies have therefore investigated the evolution of the UV slope across cosmic time until the Epoch of Reioinzation (EoR) \citep[e.g.,][]{topping22,tacchella22,robertsborsani22,cullen23,austin24}. 
Taking their results together, and combining them with pre-JWST studies (e.g., \citealt{finkelstein12,hathi13, kurczynski14,bouwens14, hathi16, pilo19, morales24}), they overall suggest that $\beta$ decreases on average from cosmic noon ($z \simeq 2$) to dawn ($z \simeq10$) for galaxies with similar stellar masses. 
In addition, several works have found that  the average UV slope of faint star-forming galaxies ($-20 < $ M$_{\rm UV}$ $<-17$) at $z\simeq 9.5$ approaches values of -2.5, i.e. close to the theoretical lower limit of $\simeq-2.6$ produced by pure stellar and nebular continuum emission \citep{topping22}.
Some studies have even found galaxies with extremely low $\beta$ values dropping below $-3$ (e.g., \citealt{topping22,atek23, austin23}, Yanagisawa et al. 2024), even though these results may be affected by photometric uncertainties and by the limited number of photometric bands used for the $\beta$ estimation. Such extremely blue slopes could be explained only if nebular emission is suppressed i.e. by the possible leakage of ionizing radiation \citep{messa24}.

Meanwhile, spectroscopic observations with JWST NIRSpec \citep{jakobsen22} have enabled the confirmation of an increasing number of galaxies in the EoR. These observations have paved the way for a statistical investigation of their physical properties from emission lines, including star formation rates \citep{calabro24}, ionization properties \citep{reddy23,sanders23}, and ISM metallicities \citep{nakajima23, curti23, sanders24}. The NIRSpec spectra have also allowed for the first time to derive $\beta$ estimates that are independent of those obtained from the photometry alone. 

Measuring the UV slopes from the spectra has several advantages.
First, we can precisely define the wavelength range for the fit, for example, following the original range between $1250$ and $2750$ \AA~rest-frame defined by \cite{calzetti94}, in order to exclude the Lyman break on the bluer side and the Balmer break on the redder side. 
Secondly, we can identify and mask bright UV rest-frame emission lines, which might cause systematic deviations from the true UV slope of the stellar continuum up to $0.5$-$0.6$, as shown by \cite{austin24}.
Finally, we can check the spectra for the presence of Lyman-$\alpha$ damping wing absorptions redward of Ly-$\alpha$, which can yield a redder UV slope if the bluest photometric band includes a substantial region close to the Ly-$\alpha$ line. 

In this work, we assemble a large spectroscopic sample of galaxies at $z > 4$ to study the evolution of the spectroscopic $\beta$ during and shortly after the end of the reionization epoch, and compare the results with previous studies based on photometry alone. To this aim, we put together JWST NIRSpec observations available from multiple surveys, including the CEERS early release survey \citep{finkelstein23}, the JADES survey \citep{eisenstein23}, and all other public surveys whose spectra have been reduced and made available through the DAWN JWST Archive 
(DJA, \citealt{heintz24}). We consider only the spectra taken with the PRISM configuration, which provides, among all spectroscopic setups, the highest S/N on the continuum, hence it is the best suited for the goals of this work.  

The paper is structured as follows: in Section \ref{sec:methodology}, we describe the observations and methodology, including the sample selection and the measurement of the UV slope $\beta$ from the PRISM spectra. In Section \ref{sec:results}, we compare the UV slopes to the UV rest-frame magnitudes of all the galaxies in the sample and study the evolution of $\beta$ with redshift. We also check for evidence of Lyman-$\alpha$ damping wing absorption in the spectra, which would be an indication of increasing IGM opacity to Ly$\alpha$ at higher redshifts. In Section \ref{sec:extremely_blue}, we identify a subset of extremely blue galaxies with $\beta < -2.6$, and we investigate the physical origin of their extreme UV slopes by comparing them to a redshift and M$_{\rm UV}$ matched sample of redder galaxies. Finally, we summarize our findings in Section \ref{sec:summary}. 
Throughout the paper, we adopt a Chabrier 2003 initial mass function (IMF), a solar metallicity of $12$ + log(O/H) $= 8.69$ \citep{asplund09}, and we assume a standard cosmology with $H_{0}=70$ $\rm km\ s^{-1}Mpc^{-1}$, $\Omega_{\rm m} = 0.3$, and $\Omega_\Lambda = 0.7$.

\section{Methodology}\label{sec:methodology}

In this study we consider JWST NIRSpec observations available from a series of programs, and select galaxies with a redshift above $4$ in order to cover the entire rest-frame UV range needed for the analysis of the UV slope. This also allows us to trace the evolution of the $\beta$ as we approach the reionization epoch. We consider all observations carried out with the PRISM/CLEAR NIRSpec configuration given the wider wavelength coverage compared to medium and high resolution NIRSpec gratings (from $0.6$ to $5.3 \mu$m), and the significantly higher sensitivity to the continuum, thus maximizing the number of sources where it is possible to measure the UV spectral slope. In the following subsections, we describe the spectroscopic observations used for the analysis, the measurement of the UV slopes, and the final sample selection. 

\bgroup
\renewcommand{\arraystretch}{1.2}
\begin{table*}[]
\centering
\caption{Contributions of each program used in this work to the final total sample.}
\label{tab:sample}
\resizebox{.6\textwidth}{!}{%
\begin{tabular}{|lll|c|c|}
\hline
\multicolumn{1}{|c|}{\textbf{DATABASE}} & \multicolumn{1}{c|}{\textbf{PROG ID}} & \multicolumn{1}{c|}{\textbf{FIELD}} & \textbf{ GAL} & \textbf{SEL } \\ \hline
\multicolumn{1}{|l|}{CEERS} & \multicolumn{1}{l|}{ERS 1345} & EGS & 109 & 79 \\ \hline
\multicolumn{1}{|l|}{\multirow{6}{*}{JADES}} & \multicolumn{1}{l|}{\multirow{6}{*}{\begin{tabular}[c]{@{}l@{}}GTO 1180\\ GTO 1181\\ GTO 1210\\ GTO 1286\\ GTO 1287\end{tabular}}} & \multirow{3}{*}{GOODS - North} & \multirow{3}{*}{249} & \multirow{3}{*}{193} \\
\multicolumn{1}{|l|}{} & \multicolumn{1}{l|}{} &  &  &  \\
\multicolumn{1}{|l|}{} & \multicolumn{1}{l|}{} &  &  &  \\ \cline{3-5} 
\multicolumn{1}{|l|}{} & \multicolumn{1}{l|}{} & \multirow{3}{*}{GOODS - South} & \multirow{3}{*}{373} & \multirow{3}{*}{225} \\
\multicolumn{1}{|l|}{} & \multicolumn{1}{l|}{} &  &  &  \\
\multicolumn{1}{|l|}{} & \multicolumn{1}{l|}{} &  &  &  \\ \hline
\multicolumn{1}{|l|}{\multirow{10}{*}{DJA}} & \multicolumn{1}{l|}{GO 1747} & BoRG-JWST & 14 & 12 \\ \cline{2-5} 
\multicolumn{1}{|l|}{} & \multicolumn{1}{l|}{GO 1727} & COSMOS & 60 & 45 \\ \cline{2-5} 
\multicolumn{1}{|l|}{} & \multicolumn{1}{l|}{GO 2565} & COSMOS, EGS, UDS & 29 & 22 \\ \cline{2-5} 
\multicolumn{1}{|l|}{} & \multicolumn{1}{l|}{\multirow{2}{*}{\begin{tabular}[c]{@{}l@{}}GTO 1211\\ GTO 1181\end{tabular}}} & \multirow{2}{*}{GOODS - North} & \multirow{2}{*}{145} & \multirow{2}{*}{66} \\
\multicolumn{1}{|l|}{} & \multicolumn{1}{l|}{} &  &  &  \\ \cline{2-5} 
\multicolumn{1}{|l|}{} & \multicolumn{1}{l|}{\multirow{5}{*}{\begin{tabular}[c]{@{}l@{}}GO 2198\\ GO 3215\\ GTO 1210\\ GTO 1180\\ DD 6541\end{tabular}}} & \multirow{5}{*}{GOODS - South} & \multirow{5}{*}{117} & \multirow{5}{*}{91} \\
\multicolumn{1}{|l|}{} & \multicolumn{1}{l|}{} &  &  &  \\
\multicolumn{1}{|l|}{} & \multicolumn{1}{l|}{} &  &  &  \\
\multicolumn{1}{|l|}{} & \multicolumn{1}{l|}{} &  &  &  \\
\multicolumn{1}{|l|}{} & \multicolumn{1}{l|}{} &  &  &  \\ \hline
\multicolumn{3}{|c|}{\textbf{TOTAL}} & \textbf{1096} & \textbf{733} \\ \hline
\end{tabular}%
}
\tablefoot{Programs used in our work and their contribution to the final sample. The column GAL include all sources in the spectroscopic databases residing in non-magnified fields after having removed the 26 identified AGN, while the SEL column reports the number of sources after implementing the additional condition on S/N $>3$.}
\end{table*}

\subsection{Spectroscopic observations }\label{sec:spec_observations}
\egroup

We first collected NIRSpec-PRISM spectra available from the Cosmic Evolution Early Release Science survey (CEERS; ERS 1345, PI: Finkelstein) in the Extended Groth Strip (EGS) field of CANDELS \citep{grogin11, koekemoer11}. In the same field, we considered sources from the DDT program 2750 (PI: Arrabal Haro). 
We then added sources from the JWST Advanced Deep Extragalactic Survey \citep[JADES,][]{eisenstein23} in the Great Observatories Origins Deep Survey \citep[GOODS,][]{giavalisco04} South field and in the GOODS North field. The fully calibrated spectra are publicly available through their third data release \citep{d'eugenio24}\footnote{\url{https://jades-survey.github.io/scientists/data.html}}. 
Finally we added galaxies in the GOODS (North and South), COSMOS, EGS, and UDS fields taken from the DAWN JWST Archive (DJA\footnote{\url{https://dawn-cph.github.io/dja/index.html}}, \citealt{heintz24}), an online repository containing reduced images, photometric catalogs, and spectroscopic data from all public JWST data products. This repository also includes galaxies observed by the BoRG-JWST survey \citep[P.I. Roberts-Borsani, ][]{robertsborsani24borg}. The DJA spectroscopic archive (DJA-Spec) is, at the time of writing, comprised of observations taken from some of the large Early Release Science (ERS), General Observer (GO) and Guaranteed Time (GTO) Cycle 1 \& 2 programs. All data processing is performed with the publicly available \texttt{Grizli} \citep{brammer23} and \texttt{MSAExp} \citep{brammer23} software modules. We considered the DJA NIRSpec MSA Extractions version 2. 
For all galaxies, slit loss corrections are taken into account by default by the jwst pipeline in the reduction process. We did not apply any further photometric corrections to the galaxy spectra. Indeed, we have checked in previous works that residual correction factors are very modest and not significantly dependent on wavelength in the range $0.6$-$3\mu m$ that is used for the derivation of the UV slopes \citep[Llerena et al. 2024 submitted,][]{calabro24,robertsborsani24}.

In our analysis we have excluded galaxies residing in fields affected by gravitational lensing of foreground clusters, to avoid considering additional uncertainties arising from magnification. In particular, we have excluded galaxies located behind the cluster Abell 2744 observed by the GLASS program \citep{treu23}, the RX J2129 galaxy-cluster field observed by the program DD 2767 (PI P. Kelly), and the JWST program GO 1433 (PI D. Coe) targeting the $z = 11.1$ MACS0647-JD galaxy.

\subsection{Spectral slope evaluation}\label{sec:slope_evaluation}

The rest-frame UV continuum slope ($\beta$) for each galaxy can be represented by a power-law of the type $f_\lambda \propto \lambda^\beta$, as modelled by \cite{calzetti94}. For the calculation of $\beta$, we considered the rest-frame wavelength range $1270\AA < \lambda < 2600\AA$ to include all the fitting windows in the original definition of \cite{calzetti94}. 

Since this wavelength range may contain several UV emission lines, we tested three different fitting methods to exclude the emission and absorption features that could systematically bias the UV slope estimates. 
In the first method, we fitted only the spectral data inside the $10$ windows defined by \cite{calzetti94}, which were selected to avoid the presence of the most relevant stellar and interstellar absorption features, including the $2175\AA$ dust feature. In the second method, we used the same 10 windows, but fitted only the mean value points inside each of them. Finally, we defined custom fitting windows in the UV range, using the entire wavelength range $1270\AA < \lambda < 2600\AA$ except four narrow bands ($20\AA$ wide), centered around these four main emission lines that could be present in some galaxies in our sample: Si\textsc{iv}]+O\textsc{iv}] ($1400\AA$), [C\textsc{iv}] ($1550\AA$), He\textsc{ii}+O\textsc{iii}] ($1640\AA$-$1666\AA$), and C\textsc{iii}] ($1909\AA$).  
By comparing the results obtained from these 3 methods, we found that the second method has a scatter of values significantly higher compared to the other two, while the first and the third methods give very similar results in terms of median value and scatter. We thus decided to follow the first approach in the rest of this paper. 

We then fitted the spectral range defined above with a linear relation of the type $\log f_\lambda = \beta \log \lambda + q$, using the \texttt{SciPy} function \texttt{scipy.optimize.curve\_fit}. This function computes optimal values for the parameters $\beta$ (and q) and their corresponding uncertainties through a non-linear least squares optimization technique, minimizing the sum of the squared residuals.
We also tested a Monte Carlo fit technique, performing a linear fit to $500$ spectral realizations with randomly perturbed fluxes and taking the median and standard deviation of all the $500$ measured $\beta$ values. This technique produced results consistent with the first, more stable approach. Our results and conclusions are thus robust against the specific fitting ranges and fitting methods considered. 
For a subset of sources from the CEERS survey for which we also had a UV-slope evaluation from photometric data from our previous work \citep{mascia24}, we compared the spectroscopic evaluation finding a general good agreement although with a large scatter.

We derived the absolute magnitude M$_{\rm 1600}$ of the galaxies in the sample, by evaluating the best-fit linear relation at $1600$ \AA\ rest-frame and applying the appropriate K-correction at the redshift of the galaxy. 

\subsection{Final sample selection}\label{sec:final_sample_selection}

To assemble the final sample, we required additional criteria on the quality of their $\beta$ slope measurements, and on the ionization source of the galaxies (AGN vs. SF), as explained below. 

Since we are interested in the spectral properties of star-forming galaxies, we first identified and removed from our sample all the sources with evidence of AGN emission. To this aim, we have followed the procedure outlined in \cite{robertsborsani24}, which consists on first removing  from the sample all the sources identified as  AGN in the literature (i.e., \citealt{harikane23, larson23, greene23, goulding23, kocevski23}), and then evaluating  via visual inspection if a two-component (i.e., broad plus narrow) model was clearly required to reproduce the \Ha\ or \Hb\ luminosity profile when the \xOIII line was well-fit by a single, narrow component (characteristic of Narrow line AGN, NLAGN). 
Morevoer, we removed AGN identified in \cite{mazzolari24} and \cite{scholtz23} with NV$1240$ \AA\ emission, which, given its ionization potential of $77.4$ eV, is very unlikely to be produced even in most extreme star-forming sources at high redshift. Finally, we removed  1 AGN listed in \cite{taylor24} that detected 50 H$\alpha$ broad-line AGN at redshifts $3.5 < z < 6.8$ using data from the CEERS and RUBIES surveys.    

In the end  we  removed from our sample a total of 26 AGN. Naturally, this approach does not rule out the possibility of further AGN contamination in our sample. The limitation in the S/N and the resolution of our spectra, and the lack of reliable identification methods at very high redshift (especially for narrow-line AGN) hampers a more accurate assessment of the AGN fraction in our sample. We also note that typical diagnostic diagrams based on UV/optical lines used in the low redshift Universe are of difficult interpretation and become less reliable at $z > 3$ \citep{calabro23, kewley19}, as the lower metallicity of AGN and the enhanced ionization parameter of galaxies in the early Universe produce line ratios that do not allow us to easily discriminate star-forming galaxies from AGN. 

The parent sample of  $1096$ galaxies in the redshift range $z > 4$ residing in non-magnified fields is presented in Table \ref{tab:sample} (column "GAL"), where we have specified the ID of the JWST program that obtained the NIRSpec-PRISM spectra and the field observed. We then 
 identified by eye and removed poor quality fits as due to a poorly detected continuum, or to bad spectral features, noise spikes, or missing spectral regions in the reduction process. All these effects could lead to unrealistic values of $\beta$ and to unreliable measurements.  
We found that imposing a minimum S/N per pixel of $3$ (averaged over the same UV continuum window used for the spectral slope estimation) yields a good threshold to discriminate between spectra with a physically reasonable value of $\beta$ (between $-4$ and $0$) and those that are too noisy to be fitted correctly.

Applying the cut in S/N $\geq 3$ leaves us with a final sample of $733$  galaxies, which are reported in Table \ref{tab:sample} (column "SEL") separately for each program  and field.

\section{\texorpdfstring{Results: UV-$\beta$ evolution}{Results: UV-beta evolution}}\label{sec:results}


\subsection{\texorpdfstring{$\beta$ vs. $M_{\rm UV}$}{beta vs. muv}}

We find UV $\beta$ slopes ranging between $-3.7$ and $0$ (median value $-2.2$ with $1 \sigma$ dispersion of $0.6$) in the entire redshift range analyzed, while M$_{\rm UV}$ values range between $-21$ and $-15.5$, with a median value of $-19$. 

\begin{figure*}[t!]
     \centering
     \includegraphics[angle=0,width=0.85\linewidth,trim={0cm 0cm 0cm 0cm},clip]{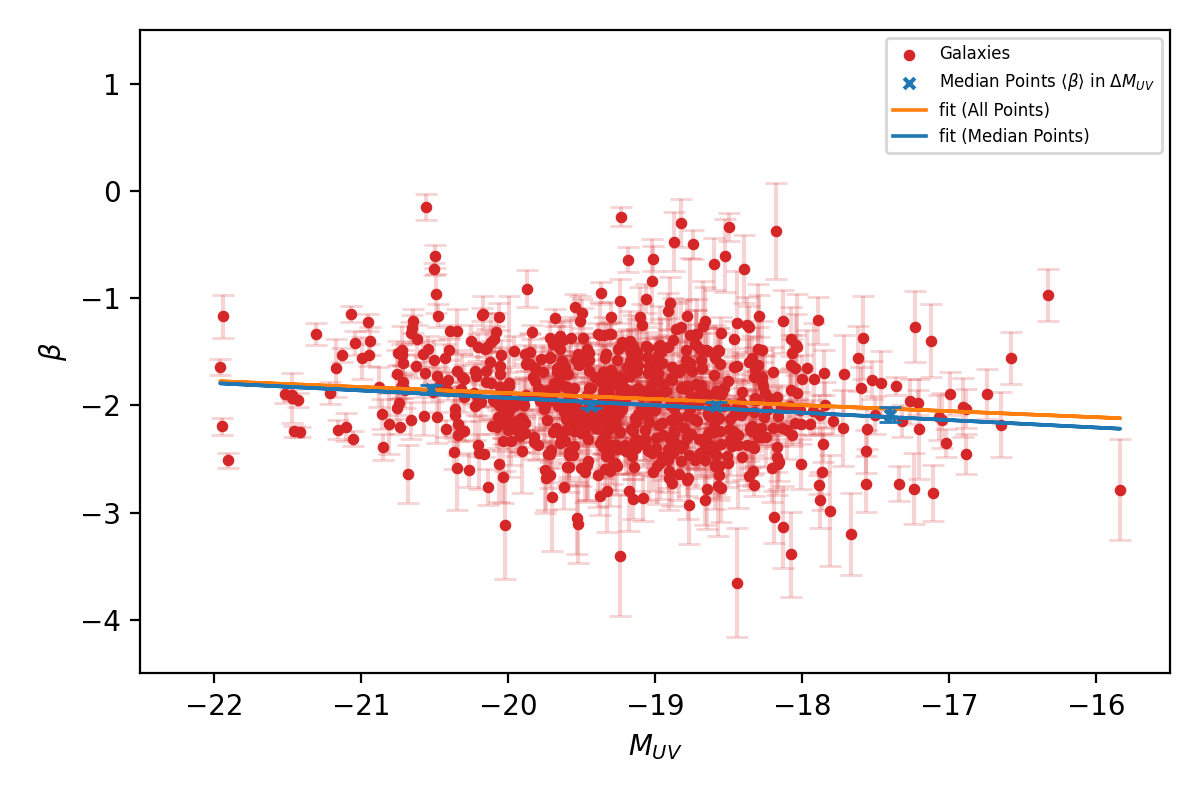}
     \vspace{-0.2cm}
     \caption{Figure showing the spectroscopic $\beta$ as a function of the absolute magnitude M$_{\rm UV}$ for the entire sample (red points) in the redshift range between $z=4$ and $z=14$. The best-fit linear relation of all the individual galaxies is shown with a orange continuous line, while the best-fit to the median points in bins of M$_{\rm UV}$ (blue crosses) is displayed with a blue continuous line.  
     }\label{Fig:beta_MUV}
\end{figure*}

\begin{figure}[h!]
    \centering
    \includegraphics[angle=0,width=1\linewidth,trim={0.02cm 0.02cm 0.02cm 0.02cm},clip]{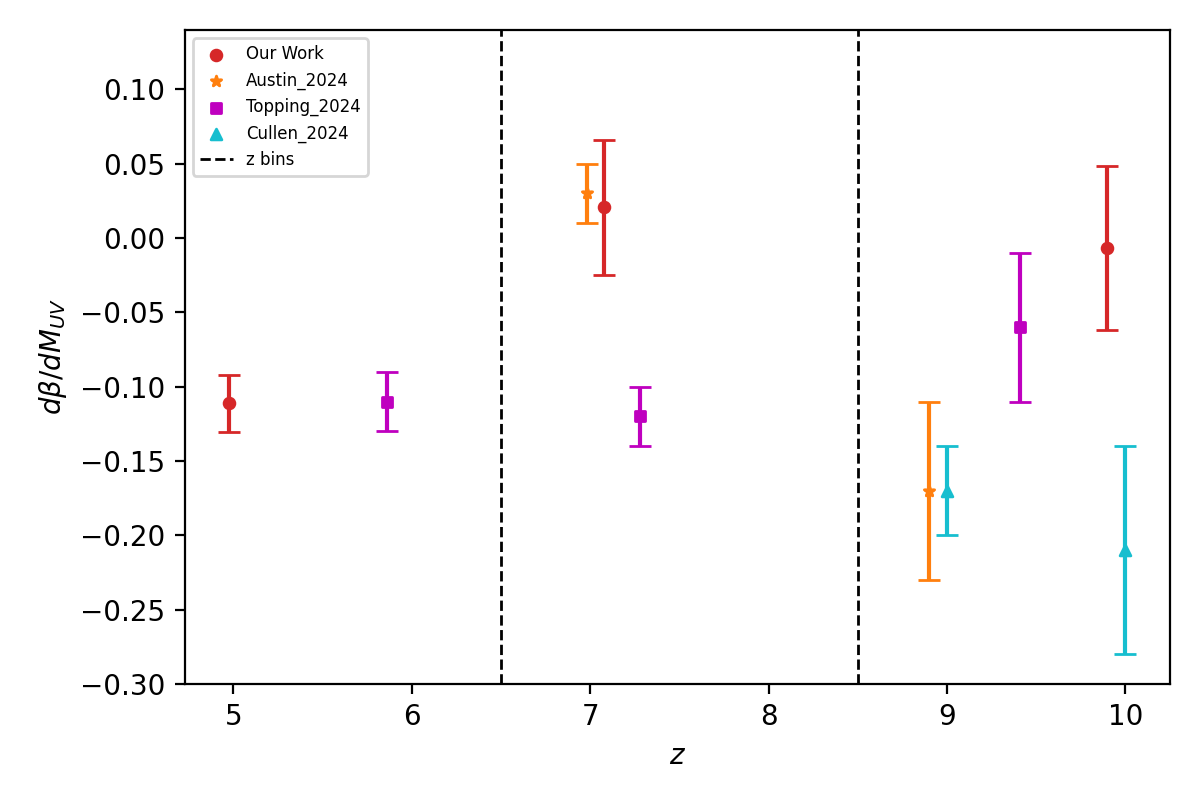}
    \caption{Figure showing the slope of the $\beta$ vs. M$_{UV}$ relation in three different redshift bins spanning the range of our analysis. }\label{z_evolution_beta_MUV}
    \vspace{-0.3cm}
\end{figure}

In Figure \ref{Fig:beta_MUV} we present the $\beta$-M$_{UV}$ relation for our sample. Considering all the data points, we derive the best fit relation:
\begin{equation}
    \beta = (-0.056 \pm 0.017) \times M_{\rm UV} - (3.01 \pm 0.34).
\end{equation}
A similar result is obtained when fitting a linear relation to the median $\beta$ in different bins of M$_{\rm UV}$. 
The negative, even though shallow, slope of this relation is likely the result of the lower dust content and younger ages of intrinsically fainter galaxies, as already noted in many previous studies (e.g., \citealt{bouwens12}), and might also reflect the well-established mass - metallicity relation observed across a broad range of redshifts (e.g., \citealt{tremonti04, maiolino08}). 
For example a similar slope of $-0.07 \pm 0.03$ was found by \cite{calabro21}, who derived $\beta$ from the photometry available in the same wavelength range of this study for $572$ galaxies at lower redshifts ($2<z<5$) and higher UV luminosities ($-22 < $M$_{\rm UV}$ $<-19.5$), observed by the VANDELS spectroscopic survey.

To compare our results with recent JWST studies, we further analyze  the  $\delta \beta$/$\delta$M$_{UV}$ in redshift intervals.  We derive  a  more negative slope at $ z<6.5 $ ($-0.11 \pm 0.02$ ), and a slope consistent with $0$ in the two intervals $6.5< z < 8.5$ and $z>8.5$ as shown in Fig. \ref{z_evolution_beta_MUV}.
This would indicate a flattening of the $\beta$ vs. M$_{UV}$ relation at higher redshifts.
A possible reason for this might be the absence from the higher redshift samples of faint sources (approximately fainter than -18.5) due to the fact that our sample is purely spectroscopically selected and does not include observations of lensed fields. However a similar outcome was suggested by \cite{topping24} who used a photometric  sample and also find a flattening of the slope when going to higher redshifts, with a $\delta \beta$/$\delta$M$_{UV}$ consistent with our result at $8.5<z<12$. We must note that,  while we agree with their median value at the low and high redshift end, our relation is flat also in the intermediate redshift range (approximately $6.5 < 8.5$) where they instead find a negative slope, although the redshift ranges considered  in the two studies  are not identical due to a different selection.
We  find an opposite trend than that reported by \cite{austin24}, with  their 
 $\delta \beta$/$\delta$M$_{UV}$ relation being flat in the intermediate redshift range (in agreement with ours) but becoming substantially steeper at $z>8.5$. Finally, \cite{cullen24} also find such steep negative relations in the very high redshift end ($z>9$). 
 
Overall, Fig. \ref{z_evolution_beta_MUV} suggests that there is a large scatter in the derived  $\delta \beta$/$\delta$M$_{UV}$ and it is still difficult to find an agreement among different studies,  likely due to combination of different methods (spectroscopic slopes vs. photometric ones), sample selection and wavelength range employed for the derivation of the  $\beta$ slope.

\subsection{Redshift evolution}

\begin{figure*}[t!]
     \centering
     \includegraphics[angle=0,width=0.85\linewidth,trim={0cm 0cm 0cm 0cm},clip]{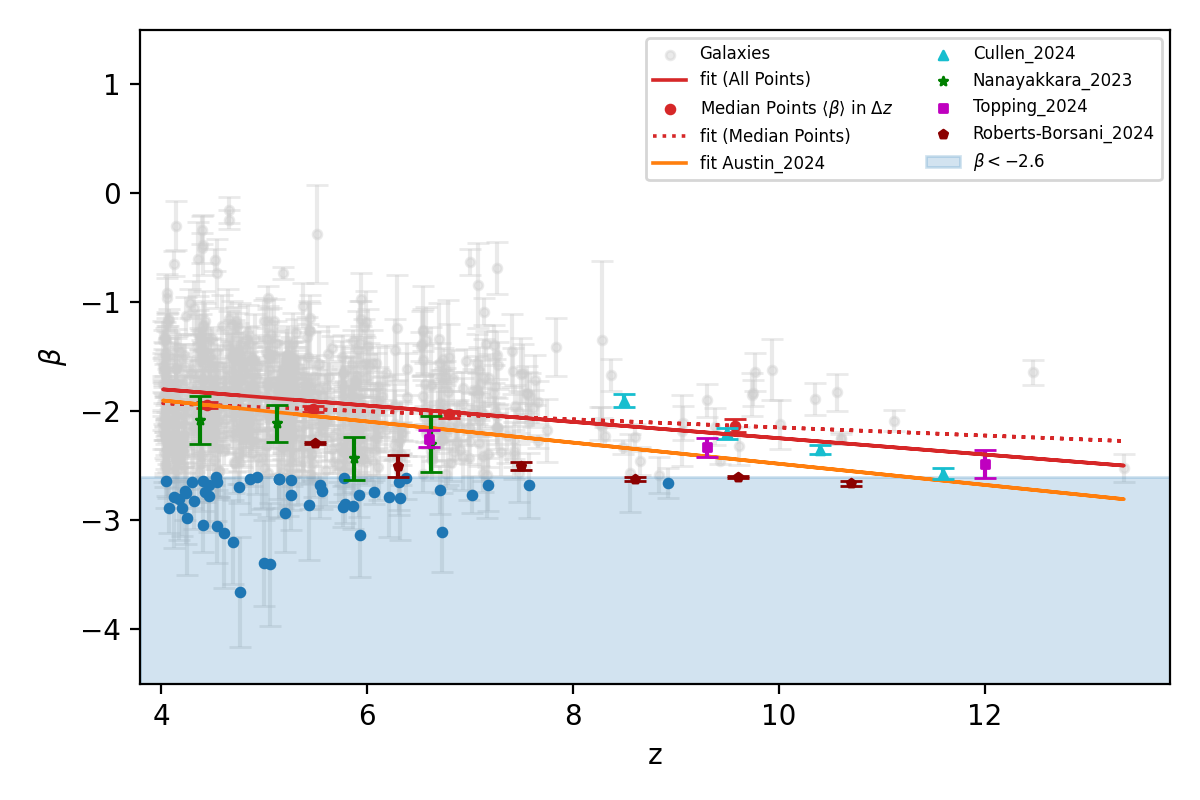}
     \vspace{-0.2cm}
     \caption{Figure showing the redshift evolution of the spectroscopic $\beta$ from $z=4$ to $z=14$. The best-fit linear relation of all the individual galaxies (grey points) is shown with a red continuous line, while the best-fit to the medium points in bins of redshift (red circles) is displayed with a red dashed line. Average UV slopes from the literature are also shown for comparison. The blue points are the extremely blue galaxies studied in Section \ref{sec:extremely_blue}.
     }\label{Fig:beta_redshift}
\end{figure*}

We then investigate how the UV slopes of galaxies evolve at $z>4$. 
To this aim, we plot $\beta$ as a function of the spectroscopic redshift for the entire sample, and the results are shown in Fig. \ref{Fig:beta_redshift}. We can see that the UV slope of galaxies becomes bluer on average from $z=4$ to $z\simeq10$, with median $\beta$ values decreasing by $0.3$ from $-1.9$ at $z\sim4.5$ to $-2.1$ at $z\sim10$. A global linear fit yields : 
\begin{equation}\label{Eq:beta_z_bestfit1}
\beta = (-0.075 \pm 0.010) \times z - (1.50 \pm 0.06).
\end{equation}
We also performed a  linear fit to the median $\beta$ values  calculated in the following redshift bins: $4$-$5$, $5$-$6$, $6$-$8$, $8$-$14$, yielding a slightly shallower relation as:
\begin{equation}\label{Eq:beta_z_bestfit2}
\beta = (-0.0371 \pm 0.0003) \times z - (1.776 \pm 0.002).
\end{equation}
A similar slope and intercept are obtained when considering the UV slopes estimated with the Monte Carlo fit method. An evolution to bluer UV colours at higher redshifts is not only of physical significance but also expected as galaxies become progressively metal and dust-poor at earlier cosmic times. Prior to JWST, this trend had been observed up to $z \approx 8$ (e.g., \citealt{bouwens14}).

The slope of our $\beta$-$z$ relation is consistent with most previous results based on the photometric UV slopes, including \cite{austin24} and \cite{robertsborsani24}. The consistency is slightly better for the steeper linear fit obtained considering all the individual points. 
We obtain a normalization of the relation that is consistent with that obtained by \cite{austin24} within $1\sigma$ uncertainty, with a small positive offset ($\sim +0.1$) emerging toward higher redshifts, where we have poorer statistics. 
This overall consistency suggests that no significant biases exist between spectroscopic and photometric-based UV slope estimates once proper correction factors for emission lines are taken into account in the latter method. 
A larger offset of $\sim +0.5$ is found instead with the relation of \cite{robertsborsani24}, with our slopes systematically redder. This is likely due to the different wavelength range ($1600 < \lambda_{\rm rest} < 2800$) used for fitting the UV slope in their work. Including the bluest portion of our fitted range as we do,  might contribute to slightly redden the global shape of the UV continuum as we will further discuss in Section \ref{sec:DLA}. 

In the redshift range between $4$ and $7$, we obtain slightly redder UV slopes on average (by $\sim0.2$) compared to \cite{nanayakkara23}, even though they are consistent within the uncertainties, and we find a similar slope of the $\beta$-$z$ relation. This work also uses only photometric data to compute the UV slope from  the best-fit EAzY SEDs, in the $1400$-$2000$ \AA\ range, so again redder than our range. 
Finally, we compare our results to the analysis of \cite{cullen24}, which is focused on the highest redshift range ($z>8$) and also employs photometric data. 
However, they include IGM  modeling in their derivation as they also use filters encompassing the Lyman break. 
Our median $\beta$ are consistent with their results only at $z \sim 9.5-10.5$, while being bluer at $z=8.5$ and redder at $z>11$: indeed their overall trend suggests a faster evolution of $\beta$ with redshift. The limited redshift coverage of these latter two studies does not allow us to make a global comparison from $z=4$ to $z\sim 10$. 

We have also checked for possible variations of the slope of the $\beta$-$z$ relation as a function of the UV magnitude of galaxies, as suggested by some recent works [ref.]. For this reason, we have split our sample into three different bins of M$_{\rm UV}$ (M$_{\rm UV}$ $< -19$, $-19 <$ M$_{\rm UV}$ $<-18$, $-18 <$ M$_{\rm UV}$ $<-17$). We find no significant variations in the slope of the best-fit relation across the entire redshift range in galaxies down to $M_{\rm UV}=-18$,  in agreement e.g. with the result found by \cite{austin24}.
We are not able to constrain the full redshift evolution for the faintest magnitude  bin since for reason already stated in the previous section we essentially lack sources with  $M_{\rm UV}$ fainter than -18 at z>8 .

To conclude, our analysis suggests that the spectroscopic UV $\beta$ follows a mild but significant decrease at higher redshifts, which is similar to the result obtained with the photometric $\beta$. This trend is consistent with the scenario where galaxies become increasingly more metal- and dust-poor toward earlier cosmic epochs, as already discussed in several pre-JWST studies (e.g., \citealt{dunlop13, finkelstein12, bouwens12, wilkins16}).

\subsection{\texorpdfstring{Evidence for Lyman $\alpha$ Damping Wing}{Evidence for Lyman alpha damping wing}}\label{sec:DLA}

\begin{figure*}[t!]
     \centering
     \includegraphics[angle=0,width=0.8\linewidth,trim={0cm 0cm 0cm 0cm},clip]{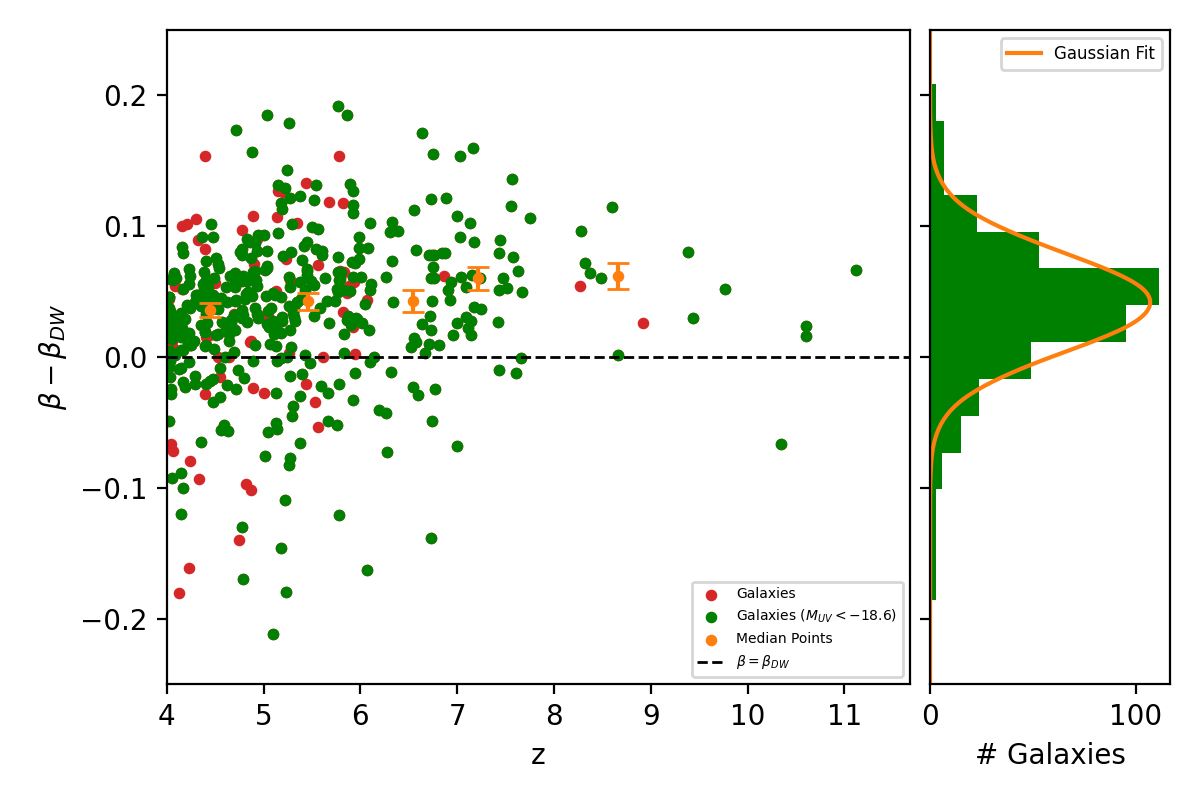}
     \vspace{-0.2cm}
     \caption{Figure showing the redshift evolution of the function $\beta - \beta_{DW}$. All galaxies reported in the plot have $S/N > 5$. Furthermore, the 407 green galaxies shown have additionally the condition $M_{UV} < -18.6$. The orange points represents the median values of the green points in redshift bins. On the right panel, an histogram of the green points is shown, fitted also by a Gaussian function with by  $\mu = 0.042 \pm 0.002$ and  $\sigma = 0.038 \pm 0.001$
     }\label{Fig:DLA}
\end{figure*}

During the EoR, Ly$\alpha$ photons from the galaxies are absorbed by neutral hydrogen present in the surrounding intergalactic medium (IGM). At high redshift, the line is so saturated that even photons emitted redward of the Ly$\alpha$ resonance can suffer significant absorption from the strong damping wings of that transition, resulting in a characteristic shape of the spectrum immediately bluewards of the Ly$\alpha$ emission and up to 100 \AA\ from resonance. 
The most widely used prescription of the Ly$\alpha$ Damping Wing (DW) is presented in \cite{miraldaescude98}. This model assumes that the IGM has a constant (volume-averaged) neutral hydrogen fraction, $x_{HI}$, between the source redshift, $z_{gal}$, and that at the end of reionization, $z_{Re}$. Recent comparisons to theorized damping wing profiles assume instead a more realistic patchy reionization process (e.g., \citealt{keating24}).

Before JWST, DW features in the EoR were typically searched for in bright QSO spectra, as their intense flux allows us to obtain high S/N spectra \citep{mortlock11, banados18}. NIRSpec-PRISM observations are finally allowing us to also obtain comparable high S/N spectra for galaxies, which are much more numerous and thus probe less biased regions of the Universe: several works have attempted measurements of Ly$\alpha$ DW to constrain the reionization process (e.g., \citealt{umeda24, witstok24}).
However, the additional effect of Damped Lyman $\alpha$ (DLA) systems with high column densities of neutral hydrogen $N_{HI} > 2 \cdot 10^{20} cm^{-2}$ \citep{lanzetta00}, arising from dense gas clouds associated with the galaxies  must also be considered. DLA have been observed at high-redshift both in quasar \citep{totani06} and galaxy spectra \citep{heintz24}: this component should be added to the IGM effect, thus complicating the derivation of the real IGM opacity (e.g., \citealt{park24}).  


The opacity due to DLA can be distinguished from the IGM effect thanks to the different wavelength dependence, but this requires high-resolution spectra that typically do not reach a high S/N in the continuum. We approach the problem by studying the average UV $\beta$ slopes of our large sample of galaxies as a function of wavelength and redshift. 
Specifically, we re-calculate the slopes excluding the region of the spectra that can potentially be affected by flux reduction due to DW effects of the IGM and/or the DLA systems, and could thus lead to redder-than-expected UV-slopes. 
We derived  $\beta_{DW}$ as the slope calculated using all the \cite{calzetti94} windows except the first two, i.e. starting at 1340 \AA\ instead of 1270.
We then focused on the difference $\beta - \beta_{DW}$; if in the range $1270\AA < \lambda < 1340\AA$ the flux is reduced because of the Ly$\alpha$ DW, $\beta_{DW}$ will consequently be bluer than $\beta$, and therefore the difference $\beta - \beta_{DW}$ will be positive.  
We note that we are assuming that the intrinsic spectra are perfectly power law in this range, which might not be the case \citep{bouwens12}. However we are specifically interested only in
studying the average $\beta - \beta_{DW}$ as a function of redshift to assess if there is a substantial evolution of this difference around the EoR, which could be probably attributed only to the increasingly neutral  IGM  rather than to other effects. 
Since the presence of DLA systems could also depend on the galaxies' brightness, we further restrict our analysis to sources with $M_{UV} < -18.6$, above which we have essentially no statistics at higher redshifts, reducing the total sample to 407 galaxies. 
 

In Fig. \ref{Fig:DLA} we show our results: we note that for most sources the $\beta - \beta_{DW}$ is positive, indicating that we obtain slightly bluer slopes if we use a longer wavelength cutoff, in agreement with the expectations and with what we discussed in the previous section when comparing our results to other analysis. 
We find  $\langle \beta - \beta_{DW} \rangle_{tot} = 0.042 \pm 0.002$ for the entire sample. 

Dividing our sample in redshift bins, between $4 < z < 6$ the median value is around  $\langle \beta - \beta_{DW} \rangle_{4 < z < 6} = 0.039 \pm 0.005$, while for $z > 7.5$ the median value is $\langle \beta - \beta_{DW} \rangle_{z > 7.5} = 0.062 \pm 0.010$. This difference is significant, and could be attributed to the increasing presence of neutral hydrogen in the IGM, affecting on average the galaxies at high redshift. We also note that all galaxies at z$>7.5$ have only positive $\beta - \beta_{DW} $.
To see if this increase around $z=7.5$ is due to a physical effect or simply  to the lower statistics at high redshift (where we only have 22 galaxies)  we perform a two-sample Kolmogorov-Smirnov test. The null hypothesis is that the distributions of $\beta - \beta_{DW} $ at low and high redshift are identical. 
We obtain a p-value of 0.044; if we choose a $2\sigma$ confidence level, a p-value less than 0.05 rejects the null hypothesis, that is, the data are not drawn from the same distribution, but they have a higher value of $\beta - \beta_{DW}$ above $z \simeq 7.5$. This could indicate that at this high redshift the extra signal in the DW is due indeed to the effect of the neutral IGM, assuming of course that the effect of the DLA systems does not evolve with redshift.
In any case the effect is very small, therefore we conclude that spectra with  higher resolution and very good S/N are needed to study this effect, as discussed e.g. by \cite{park24}.

\section{Extremely blue galaxies}\label{sec:extremely_blue}


\subsection{Selection}\label{sec:selectionXBG}

We define extremely blue galaxies (XBGs) as those systems with a UV spectral slope lower than $-2.6$. This threshold corresponds to the theoretical lower limit expected for young stellar populations in case of zero dust attenuation and the full contribution from the nebular continuum, i.e., assuming that no ionizing photons escape from the galaxy \citep{robertson10, cullen17, topping22, reddy18}. This limit is also in agreement with observations of dust-poor local star-forming galaxies \citep{chisholm22}. 
In our sample, we find $51$ XBGs with $\beta < -2.6$, with $15$ of these sources having the $\beta$ slopes lower than $-2.6$ within their $1\sigma$ uncertainties. 
We have also assembled a control sample of redder galaxies (hereafter dubbed \textit{red} galaxies) with $\beta$ between $-1.8$ and $-1.5$. We selected this subset as follows: for each XBG, we have randomly selected one red galaxy with a similar redshift (within $\pm0.5$) and a similar M$_{\rm UV}$ (within $\pm0.2$), in order to exclude biases due to $\beta$ evolving with redshift and UV magnitude. This yields a redshift- and M$_{\rm UV}$- matched subset of $51$ \textit{red} galaxies. 

\subsection{Comparing the  properties of XBGs and red galaxies through a spectral stacking analysis}\label{sec:interpretation}

\begin{figure*}[t!]
     \centering
     \includegraphics[angle=0,width=0.95\linewidth,trim={0cm 7cm 0cm 0cm},clip]{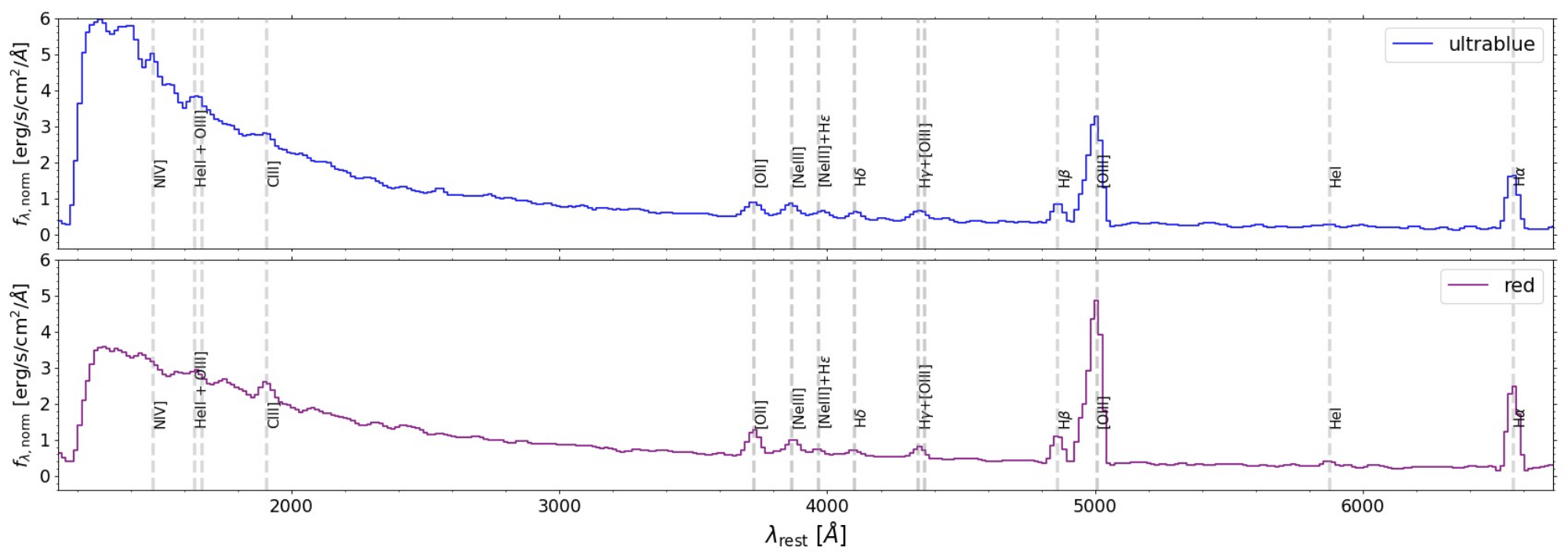}
     \vspace{-0.2cm}
     \caption{Spectral stacks of XBGs (top panel) and \textit{red} galaxies (bottom panel), between $1100$ and $6700$ \AA\ rest-frame. The position of several bright UV and optical lines is indicated with vertical gray dashed lines.
     }\label{Fig:stacks}
\end{figure*}

To investigate which mechanisms could play a role in producing very blue UV slopes, we explore how the global physical properties of XBGs differ from those of \textit{red} galaxies. Since faint optical and UV rest-frame lines (i.e., \xOII, \xNeIII, \xHeII, \xCIII) are not detected for a significant fraction of our sample, we perform a spectral stack for the subsets. 

For the stacking procedure, we adopted the same methodology used in our previous works \citep{calabro22a,calabro22b,calabro23}. In detail, we first converted all the spectra of XBGs and \textit{red} galaxies to rest-frame using their spectroscopic redshift. We then normalized them to the median flux calculated between $2500$ and $3000$ \AA\ rest-frame, and resampled them to a common grid of $2.4$ \AA\ per pixel, corresponding approximately to $1/3$ of the resolution element, from $800$ to $7000$ \AA\ rest-frame. Finally, for each pixel, we median averaged all the fluxes after a $3\sigma$ clipping rejection. The stacked noise spectrum was instead derived with a bootstrap resampling procedure with $1000$ iterations. 
The final stacks for the XBGs and the control sample are shown in Fig. \ref{Fig:stacks}, in the top and bottom panel, respectively. 

To obtain quantitative information on the nebular and stellar properties of the two populations, we measured the fluxes and the equivalent width (EW) of the most relevant emission lines detected in the entire UV+optical range, following a well-tested approach from our previous works. In brief, we fitted a gaussian profile with an underlying linear continuum in the F$_\lambda$ - wavelength parameter space, within a window of $\pm3000$ km/s from the line peak wavelength. The \xOIII+H$\beta$, \xNeIII+\xOII, and \Hg+\OIII\ in the optical, and \xHeII+\OIII\ in the UV, are fitted together. We also assumed a common redshift for the lines and a common intrinsic line width. 
From the line fluxes, we computed the following line ratio indices probing the average physical conditions in the ISM : 
\begin{itemize}
\item $\log (\text{[O {\sc ii}]} _{\lambda\lambda 3726,3729} / \text{H} \beta)$ (R2)  
\item $\log (\text{[O {\sc iii}]} _{\lambda 5007} / \text{H} \beta) $ (R3)
\item $\log \left( (\text{[O {\sc iii}]} _{\lambda\lambda 4959,5007} + \text{[O {\sc ii}]} _{\lambda\lambda 3726,3729} ) / \text{H} \beta \right)$ (R23) 
\item $\log (\text{[O {\sc iii}]} _{\lambda\lambda 4959,5007} / \text{[O {\sc ii}]} _{\lambda\lambda 3726,3729} ) $ (O32)
\item $\log (\text{[Ne {\sc iii}]} _{\lambda 3869} / \text{[O {\sc ii}]} _{\lambda\lambda 3726, 3729} )$ (Ne3O2)
\item $\log (\text{[O {\sc iii}]} _{\lambda\lambda 4959,5007} / \text{O {\sc iii}]} _{\lambda\lambda 1661, 1666} )$ (O3)
\end{itemize}

We also measured the Balmer break (hereafter also dubbed `optical break'), which is indicative of the age of underlying stellar populations. We adopted the definition of \cite{robertsborsani24}, who compare the continuum estimated in the wavelength regions between $\lambda_{\rm rest}$ $= 3500$ and $3630$ \AA\ leftward of the break, and between $= 4160$ and $4290$ \AA\ rightward of the break, to avoid contribution from strong emission lines in our low-resolution spectra, and mitigate the impact of noise. From the two continuum measurements, we finally compute the $F_{\nu,4200}/F_{\nu,3500}$ ratio. 

\begin{table}[]
\renewcommand{\arraystretch}{1.5} 
\vspace{+0.4cm}
\caption{Main spectral indices and physical properties of XBGs and red galaxies}              
\label{tab:line_ratios}      
\vspace{-0.35cm}
\begin{center} { 
\begin{tabular}{ | m{3cm} | m{2cm} | m{2.8cm} |} 
  \hline
  diagnostic & XBGs & \textit{red} galaxies \\ 
  \hline
  R2 & $-0.05 \pm 0.06$ & $0.05 \pm 0.05$ \\
  R3 & $0.78 \pm 0.06$ & $0.79 \pm 0.06$ \\
  R23 & $0.95 \pm 0.06$ & $0.97 \pm 0.05$ \\
  O3 & $1.3 \pm 0.1$ & $>1.9$ \\
  O32 & $0.96 \pm 0.03$ & $0.87 \pm 0.03$ \\
  Ne3O2 & $-0.01 \pm 0.04$ & $-0.29 \pm 0.04$ \\
  EW(HeII$1640$)  & $4.1 \pm 0.6$ & $<0.9$ \\
  EW(CIII]$1908$)  & $2.7 \pm 0.8$ & $6.3 \pm 0.4$ \\
  \large{$\frac{F_{\nu,4200}}{F_{\nu,3500}}$} & $1.04 \pm 0.01$ & $1.25 \pm 0.01$ \\
  \hline
  \hline
  physical parameter & XBGs & \textit{red} galaxies \\ 
  \hline
  $\log$(U) & $-2.12 \pm 0.03$ & $-2.20\pm0.03$ \\
  Z$_{\rm neb}$/Z$_\odot$ (strong line) &  $0.093\pm0.007$  &  $0.150\pm0.007$ \\
  Z$_{\rm neb}$/Z$_\odot$ (direct)  &  $0.08\pm0.01$  &  - \\
  $\log$ M$_\ast$/M$_\odot$ (average) & 8.0 ($\sigma =0.8$)  &  8.6 ($\sigma=0.6$) \\
  \hline
\end{tabular} }
\vspace{-0.0025cm}
\tablefoot{Table highlighting the main spectral properties (\textit{first column}) inferred from the stacked spectra of XBGs (\textit{second column}) and \textit{red} galaxies (\textit{third column}).} 
\end{center}
\vspace{-0.2cm}
\end{table}

The results of these measurements are summarized in Table \ref{tab:line_ratios} for the XBGs and the \textit{red} galaxies' stacks. 
In both cases we detect the following rest-frame optical lines: \xOII, \xNeIII, \Hg, \Hb, \xOIII, and \Ha, and \xCIII\ in the UV rest-frame. In the XBGs stack, we additionally detect the \xHeII\ and \xOIIIuv\ lines. We also tentatively detect the NIV]$1488$\AA\ line in the ultrablue stack, even though at very low significance level ($2<$ S/N $<3$), so we do not discuss it more in detail. 

Focusing on the stellar population properties, we find that the XBGs have an optical break F$_{\nu,4200}$/F$_{\nu,3500}$ that is $\sim 20\%$ smaller than in the \textit{red} galaxies sample, indicating substantially younger ages of the emitting stars. The EW of the \xHeII~line (EW $\simeq 4$\AA) detected in the stack of XBGs could also indicate the presence of Wolf-Rayet stars \citep{schaerer96, shirazi12}, massive interacting binary stars \citep{steidel16,nakajima18}, or very massive stars with masses $ > 100\ \rm M_{\odot}$ (VMS, \citealt{schaerer24})  as main ionization sources, which are expected to contribute significantly during the first $10$ Myrs of star-formation. This further supports the idea that the two populations may have substantially different stellar ages. In principle, higher resolution spectra could help to disentangle between the WR and VMS scenarios, as broad \HeII\ lines with FWHM$_{\rm velocity} > 1000$ km/s are expected in the former case. We also find that XBGs have lower stellar masses on average (average $\log$ M$_{\ast}$/M$_\odot$ $=8.0$) compared to \textit{red} galaxies by $\sim 0.6$ dex, compatible with being less evolved systems .

We also find that the XBGs have on average a higher O32 and a higher Ne3O2 index by $\sim 0.1$ dex and $\sim0.3$ dex (respectively), indicating stronger ionising conditions in these galaxies compared to \textit{red} galaxies. These values suggest that the ionization parameter of both subsets is between $\log$(U) $\sim-2$ and $-2.5$, according to the photoionization models presented in \cite{calabro23}. A more quantitative assessment using the calibration of \cite{papovich22} on the O32 index yields a $\log$(U) $=-2.12\pm0.03$ for the XBGs, $\sim0.1$ dex higher than in the \textit{red} galaxies. 
Finally, we measure a slightly higher EW(\CIII) for the \textit{red} galaxies, even though both measurements are consistent with the EW(\CIII) of normal high-z star-forming galaxies \citep[see e.g.,][]{amorin17,nakajima18,llerena22}, and with the average range measured recently with JWST data for the global population of star-forming galaxies at $z \sim 5$ \citep{robertsborsani24}. 

The line indices presented in Table \ref{tab:line_ratios} can be used to infer the gas-phase metallicity Z$_{\rm gas}$ of the galaxies. Using the full set of five spectral indices, namely R2, R3, R23, O32, and Ne3O2, we calculate Z$_{\rm gas}$ for the two spectral stacks, using the strong-line calibrations of \cite{sanders24}, built with a large subset of galaxies from $z=2$ to $z=9$ (covering a similar stellar mass and star-formation rate range of our high-redshift galaxies) and anchored to the electron temperature (T$_e$) method. 

We find that, regardless of the index, Z$_{\rm gas}$ of XBGs is always lower than Z$_{\rm gas}$ in \textit{red} galaxies. Calculating an error weighted average of the metallicities estimated with all the five indices yields Z$_{\rm gas}$ $=9.3 \pm 0.7 \%$ Z$_\odot$ in XBGs and $15 \pm 3\%$ in \textit{red} galaxies. 
In the stack of XBGs, we also detect the \xOIIIuv\ line, from which we obtain a metallicity (through the direct T$_e$ method) of $8 \pm 1 \%$ Z$_\odot$, consistent with the strong-line method.
These results indicate that there is a significant difference in metallicity, at $\sim 8 \sigma$ significance level, between the two subsets, with XBGs more metal poor compared to \textit{red} galaxies. 

The difference in metallicity between XBGs and \textit{red} galaxies reflects the strong correlation observed between the UV slope and the gas-phase and stellar metallicity in star-forming galaxies at lower redshifts \citep{shivaei20,calabro21}. Since the UV slopes are primarily regulated by the amount of dust attenuating the stellar intrinsic spectrum, this again suggests that XBGs are chemically young systems, less enriched in dust and metals compared to more evolved, \textit{red} galaxies. 

Overall, our findings suggest that XBGs have younger stellar populations, stronger ionization fields, lower dust attenuation, and lower metallicity compared to \textit{red} galaxies.  
It is worth remarking that, despite their extreme $\beta$ values, XBGs do not contain pristine gas with zero metallicity, but their ISM is already enriched at almost $10 \%$ solar. Using the SED models presented in Fig.~2 of \cite{topping22} with a stellar metallicity of $10\%$ solar, and down to a Z $= 1\%$ Z$_\odot$ as a more conservative value (to account for the possible rescaling between the gas-phase and stellar metallicity), we note that, even though stellar populations with very low metallicities and young ages have bluer intrinsic spectra, the metallicity or the age alone cannot explain $\beta$ values $< -2.6$. This analysis suggests that the only effect that can produce UV spectra significantly bluer than $\beta \simeq -2.6$ is a lower contribution of the nebular continuum emission, hence a significant leakage of ionizing radiation. We explore this scenario in the next subsection.

\subsection{The origins of the very blue UV spectral slopes}\label{physical_interpretation}

\begin{figure*}[t!]
     \centering
     \includegraphics[angle=0,width=0.999\linewidth,trim={0cm 9.4cm 7cm 0cm},clip]{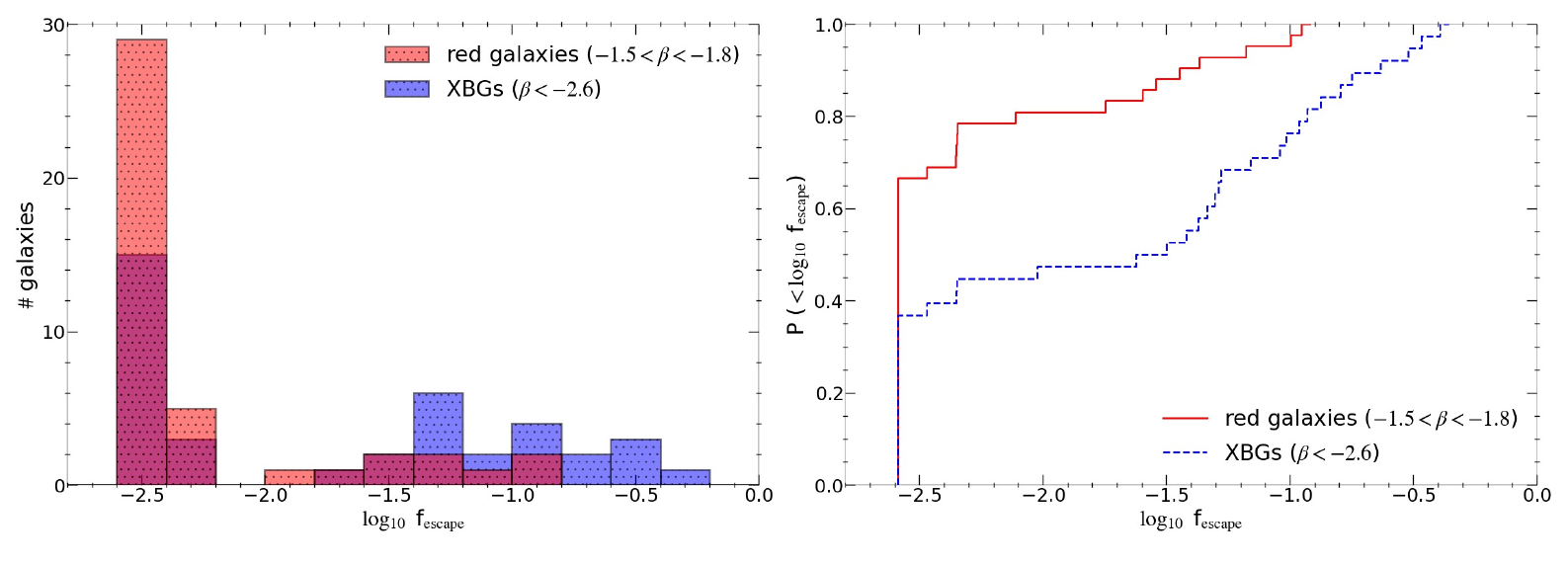}
     \vspace{-0.4cm}
     \caption{\textit{Left:} Histogram distribution of the escape fractions (estimated indirectly as explained in the text, and in logarithmic scale) for the subsets of XBGs (blue bars) and \textit{red} galaxies (red bars), which are matched in redshift and M$_{\rm UV}$ to the XBGs. \textit{Right:} Cumulative distribution functions of $\log_{10}$ \fesc\ for XBGs and \textit{red} galaxies (respectively blue dashed and red continuous line).
     }\label{fig:f_escape}
\end{figure*}

Previous studies have shown that a possible way to obtain UV slopes bluer than $-2.6$ is to reduce the contribution of the nebular continuum emission. This can happen if a fraction of the stellar continuum emission does not ionize the surrounding interstellar medium (ISM) but is free to propagate outside of the galaxy.
One way to test this scenario is calculating the escape fraction of Lyman continuum radiation (f$_{\rm esc}$), for the XBGs and the \textit{red} galaxies populations to assess if they are different.

At $z>4$ it is impossible to directly detect the Lyman continuum emission due to the IGM opacity, but several works have shown that it can be estimated f$_{\rm esc}$ in an indirect way. The simplest  method is applying directly the relation between $\beta$ and $\rm f_{esc}$ derived by \cite{chisholm22}, as previously done e.g., by \citet{cullen24}: this would would imply \fesc $> 0.19$ for all the XBGs and a median \fesc $=0.013$ for the red ones.
However this relation presents a really large scatter when applied to the Low redshift Lyman continuum survey leakers. For this reason, alternative methods based on a combination of several indirect indicators have been developed \citep{jaskot24,mascia23,choustikov24}, which give more robust estimates also for intermediate redshift leakers (\citealt{jaskot24}, Mascia et al. in prep).
We applied the Cox proportional hazards models of \cite{jaskot24}, which are based on a survival analysis technique, recalibrated by Mascia et al. (in prep) on Low redshift Lyman Continuum Survey galaxies that are analogs of EoR sources.  
In particular, we applied the `ELG-O32' model which uses a combination of the stellar mass, the stellar attenuation A$_{\rm V, stellar}$, the UV magnitude M$_{\rm UV}$, and the O32 index, defined as $\log$([O \textsc{iii}]$5007$/[O \textsc{ii}]$3727$), to indirectly infer the escape fraction of ionizing photons. The stellar mass, M$_{\rm UV}$, and the stellar attenuation A$_{\rm V, stellar}$ were derived from fitting stellar population templates with nebular contribution to the available multi-wavelength photometry from \cite{merlin24}, using the code \texttt{zphot.exe} \citep{fontana00}. 
The O32 indices were measured directly from the individual spectra, and corrected for dust attenuation using the H$\alpha$/H$\beta$ ratio, or the nebular attenuation A$_{\rm V, neb}$ $=$ A$_{\rm V, stellar}$ $/0.44$ above the redshift where H$\alpha$ is not covered  by NIRSpec. 
If [O \textsc{ii}]$3727$ is not detected, we calculated a $1 \sigma$ lower limit for f$_{\rm esc}$. 

We compare the distributions of f$_{\rm esc}$ of the XBGs and the \textit{red} galaxies populations in Fig. \ref{fig:f_escape}-\textit{left}. 
Overall, we find that both distributions are non-Gaussian, with a bulk of galaxies with very small escape fractions ($< 0.01$) and a tail of galaxies with higher f$_{\rm esc}$ similar to the results of previous works on galaxies at similar redshift \citep{mascia23,mascia24}. 
The XBGs population has an average $f_{\rm esc}=0.073$ that is $\sim 5$ times larger compared to the \textit{red} galaxies population, and it reaches a much higher maximum escape fraction ($0.44$ compared to $0.12$). A Kolmogorov-Smirnov (KS) test confirms that the XBGs and \textit{red} galaxies subsets are likely taken from two different underlying distributions (KS-statistics$=0.37$, p-value$=0.005$). 
We also compared the cumulative distribution functions (CDF) of the two XBGs and the \textit{red} galaxies (Fig. \ref{fig:f_escape}-\textit{right}). With these CDFs, we additionally performed a Mann-Whitney U test,
which returns a p-value of $5\times 10^{-4}$, again indicating that the difference in the distributions is statistically significant.

This suggests that a possible reason for the extremely blue slopes at least in some of the XBGs is the higher escape fraction of Lyman continuum radiation, also considering that in more than half of the blue sources, the inferred  f$_{\rm esc}$ are lower limits.
However, as shown by \cite{topping22}, only with extreme f$_{\rm esc}$ values,   the $\beta$  predicted by models become as blue as $-3$, whereas only a few of our XBGs have inferred f$_{\rm esc}>0.1$. 

Alternatively, the blue sources might be caught in a peculiar moment of cessation of star formation shortly after a burst, as proposed by \citet{topping24}. In this case the ionizing photon production and the resulting nebular continuum strength falls and we observe purely  stellar continuum in the UV. Since the ages are still small, the UV slope can be  very blues. Evidence for bursty star-formation histories are indeed found at the EoR from JWST studies \citep{asada24,endsley24}. We will perform a detailed spectro-photometric analysis in a future work.

\section{Summary}\label{sec:summary}
In this work we have analyzed a large sample of 733 galaxies selected from a mixture of JWST ERS/GTO/GO observational programs  with $z > 4$ and boasting spectroscopic data obtained with the low resolution ($R \sim 30-300$) PRISM/CLEAR NIRSpec configuration. We have determined their UV continuum slopes $\beta$ with the main of studying  the evolution of the average properties of galaxies over a large cosmic epoch. We also  identified  a subset of sources showing spectral slopes bluer than -2.6, i.e. bluer than the lower limit that can be reached with stellar plus nebular continuum emission, in case of no dust attenuation. We analyzed them both individually and through  spectral stacking, to uncover the physical origin of such extremely blue slopes.

We summarize the main findings of our work  as follows : 
\begin{enumerate}
    \item We find a shallow dependence of $\beta$ on $\rm M_{UV}$ (best-fit slope of $-0.056 \pm0.017$) for the entire sample. We also find that the $\beta$ - $\rm M_{UV}$ relation tends to flatten towards higher redshifts, going from a slope of $ \delta \beta /\delta M_{UV} \sim-0.1$ at $z<6.5$ to $\sim0$ at $z>6.5$.
    \item The average $\beta$ slopes follow a mild but significant evolution with redshift, with galaxies becoming bluer toward earlier cosmic epochs ($ \delta \beta /\delta z = - 0.075$). These findings are consistent with the results obtained for galaxies with similar redshifts and UV luminosities and UV slopes measured from photometric data. 
    \item We analysed the presence of the Ly$\alpha$ damping wing  in the spectra, which we know is due to a combination of the presence of increasing neutral hydrogen in the IGM as well as the DLA systems associated to the galaxies. While the first effect should only appear when the IGM is significantly neutral ($z>7$), the latter should be redshift independent in an $\rm M_{UV}$ limited sample. We find a small but significant effect on the $\beta$ slope at $z>7.5$, suggesting an increase of the neutral IGM content,  which is however difficult  to quantify with present low resolution data. 
    \item We obtained a stacked spectrum of the galaxies  with extremely blue UV slopes ($\beta <-2.6$), and compared the average properties determined from the detected emission lines to an analogous spectral stack of a sample of red sources ($-1.8<\beta<-1.5$) matched in $\rm M_{UV}$ and redshift. We found that XBGs have higher ionization parameter, lower (although not pristine) metallicity, and younger ages compared to redder galaxies. However, these properties alone cannot explain their extreme $\beta$ values.
    \item Using the most solid indirect predictors of Lyman continuum escape recently developed (\citealt{jaskot24}, Mascia et al. in prep.) we find that the escape fraction f$_{\rm esc}$ may be $>10\%$ in a large fraction of the XBGs, contrary to what inferred for red galaxies with the same $\rm M_{UV}$ and redshift. The reduction of the nebular continuum, which is the effect of a large leakage of ionizing photons, would then explain the very blue spectral slopes. 
    However some XBGs also have very small escape fraction suggesting that other physical mechanism, such as a bursty star formation history, must also play a role. 
\end{enumerate}


Future large spectroscopic surveys, such as CAPERS, will significantly increase the number of galaxies with spectroscopic follow up in the reionization epoch. This will allow us to constrain more robustly the evolution of beta with redshift, and the slope of the $\beta$ - $\rm M_{UV}$ relation, testing also the role of other physical properties in those relations. Deeper and higher resolution spectra in the UV rest-frame will instead help to better constrain the profile of the Lyman-alpha damping wing and to disentangle the different contribution due to the DLA and IGM absorption with the help of the latest models and radiative transfer cosmological simulations.

\noindent

\begin{acknowledgements}
We acknowledges support from the INAF Large Grant for Extragalactic Surveys with JWST and  from the PRIN 2022 MUR project 2022CB3PJ3 - First Light And Galaxy aSsembly (FLAGS) funded by the European Union – Next Generation EU. PS acknowledges INAF Mini Grant 2022 “The evolution of passive galaxies through cosmic time”. Part of the research activities described in this paper were carried out with the contribution of the Next Generation EU funds within the National Recovery and Resilience Plan (PNRR), Mission 4 - Education and Research, Component 2 - From Research to Business (M4C2), Investment Line 3.1 - Strengthening and creation of Research Infrastructures, Project IR0000034 – ``STILES - Strengthening the Italian Leadership in ELT and SKA''.
\end{acknowledgements}

{}




\end{document}